\documentclass[conference]{IEEEtran} 
\IEEEoverridecommandlockouts

\usepackage{amsmath,amssymb,amsfonts, amsthm}

\usepackage{graphicx}
\usepackage{textcomp}

\usepackage{xcolor}
\usepackage{bm}

\usepackage{url}
\usepackage{cite}
\usepackage{float}

\renewcommand\IEEEkeywordsname{Keywords}
\usepackage{caption}

\usepackage{comment}
\usepackage[ruled,linesnumbered]{algorithm2e}

\usepackage{float}
\usepackage{multicol}
\usepackage{color}
\newcounter{multifig}

\usepackage{thmtools,thm-restate}
\usepackage{multirow}
\usepackage{mathtools, cuted}

\usepackage{lipsum}
\usepackage{array}
\usepackage{makecell}

\usepackage[caption=false, font=footnotesize]{subfig}

\def\BibTeX{{\rm B\kern-.05em{\sc i\kern-.025em b}\kern-.08em
    T\kern-.1667em\lower.7ex\hbox{E}\kern-.125emX}}
\begin{document}

\title{Optimal Regulation of Prosumers and Consumers in Smart Energy Communities}


\author{\IEEEauthorblockN{Syed Eqbal Alam and Dhirendra Shukla\thanks{This work is supported in part by the Mitacs Accelerate program (grant number IT24468) and Gray Wolf Analytics, Fredericton, New Brunswick, Canada.}\thanks{The paper is published in the proceedings of the 2022 IEEE
International Smart Cities Conference (ISC2), September 2022, Paphos, Cyprus. It can be cited as:
S. E. Alam and D. Shukla, "Optimal Regulation of Prosumers and Consumers in Smart Energy Communities," 2022 IEEE International Smart Cities Conference (ISC2), 2022, pp. 1-7, doi: 10.1109/ISC255366.2022.9921890.}}
\IEEEauthorblockA{
J. Herbert Smith Centre for Technology Management \& Entrepreneurship \\
University of New Brunswick \\
Fredericton, New Brunswick, Canada \\
}}

\maketitle

\begin{abstract}
In smart energy communities, households of a particular geographical location make a cooperative group to achieve the community's social welfare. Prosumers are the users that both consume and produce energy. In this paper, we develop stochastic and distributed algorithms to regulate the number of consumers and the number of prosumers with heterogeneous energy sources in the smart energy community. In the community, each prosumer has one of the heterogeneous energy sources such as solar photovoltaic panels or wind turbines installed in their household. The prosumers and consumers decide in a probabilistic way when to be active. They keep their information private and do not need to share it with other prosumers or consumers in the community.
Moreover, we consider a central server that keeps track of the total number of active prosumers and consumers and sends feedback signals in the community at each time step; the prosumers and consumers use these signals to calculate their probabilistic intent. We present experimental results to check the efficacy of the algorithms. We observe that the average number of times prosumers and consumers are active reaches the optimal value over time, and the community asymptotically achieves the social optimum value. 
\end{abstract}


\section{Introduction}
The smart energy community has attracted significant interest from the research community recently. It consists of energy prosumers and consumers. {\em Energy prosumers} are the users that can consume and produce energy; for example, households connected to a power grid to consume energy and have solar photovoltaic panels on their rooftops to produce energy locally. In smart energy communities, members of a particular geographical location make a cooperative group to achieve a common goal. Moreover, in the smart energy communities, the prosumers provide the surplus produced energy to community members or a grid for monetary benefits or to store energy in a community energy storage.
Furthermore, the prosumers may also contribute surplus energy to philanthropic works and help handle social challenges such as {\em energy poverty}. In particular, the members can support specially-abled people. Note that energy poverty is the lack of sufficient energy for primary human activities \cite{Guruswamy2011}. 
The smart energy community utilizes the existing local energy infrastructures and enables community members' participation, cooperation, and coordination for the community's welfare \cite{Savelli2021}. In addition, the smart energy communities facilitate balancing local energy production and consumption and aid in balancing community members' needs and preferences fairly. They also help keep monetary benefits within the community. Furthermore, it can help in achieving cleaner and more affordable energy. Additionally, roughly speaking, smart energy communities can facilitate social interaction and build relationships among the community members that are unavailable in the traditional energy systems \cite{Savelli2021}. These potentials can help in societies' prosperity, and in becoming an energy community self-sustainable \cite{Mengelkamp2018}, \cite{Morstyn2018}, \cite{Ableitner2020}; furthermore, it can help in achieving net-zero emission targets. 

As the local weather affects the energy production from renewable sources, we can reduce the impact by using heterogeneous energy sources as considered in the paper. Moreover, one of the main challenges in realizing smart energy communities is developing mechanisms to handle the uncertainty in energy production and consumption patterns. Particularly, the problem of regulating prosumers and consumers with optimality constraints, wherein the overall cost to the energy community is minimized, is not well studied. Our work contributes toward addressing this problem.
%
We assume that members (prosumers) in the smart energy community have heterogeneous renewable energy sources, some prosumers install solar panels, and others install wind turbines in their households. The prosumers provide excess produced energy to some community members, called {\em energy consumers}. Costs are associated with the installation and transmission of energy from renewable sources. Our principal contribution in the paper is to develop distributed and stochastic algorithms for the smart energy community that regulate the number of energy consumers and the number of prosumers with heterogeneous energy sources to minimize the overall cost to the energy community. The algorithms are motivated by the ideas from adaptive control and stochastic approximation, \cite{Borkar2008}, \cite{Griggs2016}, \cite{Alam2020}. 

The developed model consists of prosumers with heterogeneous energy sources, energy consumers, and a central server. The prosumers and the consumers join the energy community at the start of the system; they decide to be active in the community based on a probabilistic rule. Each prosumer and consumer has a cost function that depends on the average number of times they actively participate in the energy community.  Furthermore, the members do not need to share their information, for example, the cost functions, the derivatives of the cost functions, or the average number of times they were active with other members. 
Furthermore, the central server keeps track of the aggregate number of active prosumers and active consumers at a time step---the central server updates and broadcasts feedback signals for prosumers and consumers in the community. Using the feedback signals, the derivatives of the cost functions, and an average of the number of times a prosumer was active, the prosumer calculates its probabilistic intent to be active or not. Similarly, the consumers calculate their probabilistic intent to be active or not. Following the algorithms, prosumers and consumers obtain optimal values empirically on long-term averages to achieve minimum overall cost to the energy community.

\section{Related work}
This section briefly presents the recent works on smart energy communities. Chen et al. proposed a peer-to-peer energy sharing platform for the dynamic networks that reduce the power losses over the network \cite{Chen2021_1}. Pena-Bello et al. \cite{Bello2022} study the dependence of prosumer preferences and the performance of the peer-to-peer solar energy communities. Mahmud et al. studied peer-to-peer energy sharing and trading in \cite{Mahmud2021}. Furthermore, Werner and co-authors studied the pricing flexibility in energy markets in \cite{Lucien2021}. Peer-to-peer energy trading fairness mechanisms are proposed in \cite{Eusung2020} that maximize the social welfare of the smart energy communities. Furthermore, Moret and Pinson in \cite{Fabio2019} propose an energy community, named as {\em energy collective}, wherein prosumers optimize their individual utilities based on ADMM; the authors also provide analysis on the community fairness. In ADMM-based solutions, the community members share their states or Lagrange multipliers with at least one neighboring member. However, in our solutions, a member does not exchange information with other members in the community but with the central server---if the member is participating at a time step or not. The central server keeps track of the aggregate number of active prosumers and consumers in the community.
In another direction, a reinforcement learning-based mechanism is proposed for peer-to-peer energy trading in smart energy communities \cite{Zhou2019}.
Bhattacharjee and Nandi developed a voting-based mechanism for supply and demand balancing in energy communities with three heterogeneous renewable energy sources---solar, wind, and biomass \cite{Bhattacharjee2021}. 
Recently, blockchain and distributed ledgers based peer-to-peer energy trading systems are studied in \cite{Wang2019}, \cite{Bajpai2021}, \cite{Tarek2022}, \cite{Matthew2022}, \cite{Zia2020}, to name a few. 
Finally, readers may refer to the survey article on smart energy communities \cite{Ceglia2020}, \cite{Tushar2021}, \cite{Esmaeil2021}, \cite{Debora2021}, \cite{Thornbush2021} and the papers cited therein. 

\section{Preliminaries}  \label{prob_form}
Consider a smart energy community with heterogeneous energy sources such as solar panels and wind turbines; in the community, energy prosumers and consumers group together to take care of the community's energy needs and minimize the overall cost to the community. Moreover, let $N$ prosumers have solar panels and $M$ prosumers have wind turbines in the community. The prosumers sell the excess produced energy to the community members or share it with needy consumers in the community as philanthropic work. Let $U$ be the number of energy consumers in the community. We assume a central server keeps track of the aggregate number of active prosumers and active consumers at a time. Let every prosumer with a solar panel have a cost function that depends on the average number of times the prosumer was active in sharing the produced energy. Analogously, every prosumer with a wind turbine has a cost function that depends on the average number of times that prosumer was active. Similarly, every consumer has a cost function that depends on the average number of times the consumer was active. Furthermore, let the total desired number (or the capacity) of prosumers with solar panels be $C_\textnormal{s}$ and the desired number of prosumers with wind turbines be $C_\textnormal{w}$ in the community. 
%

Let $\mathbb{N}$ denote the set of natural numbers, and let $k \in \mathbb{N}$ denote the time steps. Let $\xi_{i}(k) \in \{0,1\}$ denote whether prosumer $i$ with solar panel is active at time step $k$ or not. When the prosumer $i$ with solar panel is active then $\xi_{i}(k) = 1$ is updated; otherwise, $\xi_{i}(k) =0$. Let $\eta_{j}(k) \in \{0,1\}$ denote whether prosumer $j$ with wind turbine is active at time step $k$ or not. When the prosumer $j$ with wind turbines is active then $\eta_{j}(k) =1$ is updated; otherwise, $\eta_{j}(k) =0$. Furthermore, let $x_{i}(k) \in [0, 1]$ denote the average of the number of times prosumer $i$ with solar panels was active until time step $k$, define, for $i = 1, 2, \ldots, N$, as follows:
\begin{align} \label{eq:average_prod-solar}
{x}_{i}(k) \triangleq \frac{1}{k+1} \sum_{\ell=0}^k \xi_{i}(\ell), \quad i = 1, 2, \ldots, N.
\end{align}
Let $y_{j}(k) \in [0, 1]$ denote the average of the number of times prosumer $j$ with wind turbine was active until time step $k$, defined, as follows:
\begin{align} \label{eq:average_prod-wind}
{y}_{j}(k) \triangleq \frac{1}{k+1} 
\sum_{\ell=0}^k \eta_{j}(\ell), \textnormal{ for } j = 1, 2, \ldots, M.
\end{align}
Let $\zeta_{u}(k) \in \{0,1\}$ denote the energy consumer $u$ at time step $k$ was active or not. When the consumer $u$ is active then $\zeta_{u}(k) = 1$ is updated; otherwise, $\zeta_{u}(k) =0$. Furthermore, let $z_{u}(k) \in [0, 1]$ denote average of the number of times consumer $u$ was active until time step $k$. We define it as follows:
\begin{align} \label{eq:average_cons}
{z}_{u}(k) \triangleq \frac{1}{k+1}
\sum_{\ell=0}^k \zeta_{u}(\ell), \textnormal{ for } u = 1, 2, \ldots, U.
\end{align}
Note that $\xi_i(k)$, $\eta_j(k)$, and $\zeta_u(k)$ are independent Bernoulli random variables.

Let $\mathbf{x} = (x_1,x_2, \ldots, x_N) \in [0,1]^N$, and let $\mathbf{y} = (y_1,y_2, \ldots, y_M) \in [0,1]^M$, and let $\mathbf{z} = (z_1,z_2, \ldots, z_U) \in [0,1]^U$; the bold letters denote the vector entries.  Let $\mathbb{R}_+$ denote the set of positive real numbers. Let each prosumer have a cost function associated with energy production; analogously, each consumer has a cost function associated with energy consumption. More specifically, let $f_i: [0,1] \to \mathbb R_+$ be the cost function of prosumer $i$ which is associated with a cost to the production of solar energy for prosumer $i$, for $i=1,2,\ldots,N$. Furthermore, let $g_j: [0,1] \to \mathbb R_+$ be the cost function of the prosumer $j$ which is associated with a cost to production of wind energy, for $j=1,2,\ldots,M$. In addition, let $h_u: [0,1] \to \mathbb R_+$ be the cost function of consumer associated with a cost to consumption of energy, for $u=1,2,\ldots,U$. We assume that the cost functions $f_i$, $g_j$, and $h_u$ are twice continuously differentiable, convex, and increasing. Furthermore, we assume that the prosumers and consumers do not share their information with other prosumers or consumers in the community. 
Additionally, let $f_i'(x_{i})$ denote the derivative of the cost function $f_i$ of prosumer $i$ with solar panels with respect to $x_{i}$; $g_j'(y_{j})$ denote the derivative of the cost function $g_j$ of prosumer $j$ with wind turbines with respect to $y_{j}$; and $h_u'(z_{u})$ denote the derivative of the cost function $h_u$ of consumer $u$ with respect to $z_{u}$, for $i=1,2,\ldots,N$, $j=1,2,\ldots,M$, and $u=1,2,\ldots,U$.

We aim to minimize the overall cost to the energy community. We formulate the optimization problem as follows:
\begin{align}
\begin{split} \label{obj_fn1_b1}
\min_{\mathbf{x}, \mathbf{y}, \mathbf{z}} \quad &\sum_{i=1}^{N} f_i(x_{i}) + \sum_{j=1}^{M} g_j(y_{j}) + \sum_{u=1}^{U} h_u(z_{u}),
\\ \mbox{subject to } \quad & \sum_{i=1}^{N} x_{i}  =  C_\textnormal{s}, \\
& \sum_{j=1}^{M} y_{j} = C_\textnormal{w}, \\
& \sum_{u=1}^{U} z_{u} = \sum_{i=1}^{N} x_{i}  +  \sum_{j=1}^{M} y_{j},
\\  &x_{i} \geq 0,  \quad \mbox{for } i=1,\ldots,N,
\\&y_{j} \geq 0, \quad \mbox{for } j=1,\ldots,M, \\
&z_{u} \geq 0, \quad \mbox{for } u=1,\ldots,U. 
\end{split}
\end{align}

Let the optimal value of the average of the number of times a prosumer with solar panels be denoted by $\mathbf{x}^* = ({x}_{1}^{*}, \ldots, {x}_{N}^*) \in (0,1]^N$ and optimal value for prosumers with wind turbines be noted by $\mathbf{y}^* = ({y}_{1}^{*}, \ldots, {y}_{M}^*) \in (0,1]^M$, and optimal value for energy consumers be denoted by $\mathbf{z}^* = ({z}_{1}^{*}, \ldots, {z}_{U}^*) \in (0,1]^U$. The values $\mathbf{x}^*$, $\mathbf{y}^*$, and $\mathbf{z}^*$, represent the solutions to the optimization problem \eqref{obj_fn1_b1}.
In this paper, we develop distributed stochastic algorithms that determine the number of active prosumers with solar panels, the number of active prosumers with wind turbines, and energy consumers at a time step. The algorithms ensure that the average of the number of times prosumers and consumers are active, as defined in \eqref{eq:average_prod-solar}, \eqref{eq:average_prod-wind}, and \eqref{eq:average_cons}, respectively, reach optimal values asymptotically. Precisely, the number of active energy prosumers with solar panels reaches the optimal value as in \eqref{eq:avg_sol}:  
\begin{align} \label{eq:avg_sol}
\lim_{k\to \infty} {x}_{i}(k) = {x}_{i}^{*}, \text{ for } i=1,2,\ldots,N.
\end{align}
And, the number of active energy prosumers with wind turbines reaches its optimal value as in \eqref{eq:avg_wind}:
\begin{align} \label{eq:avg_wind}
\lim_{k\to \infty} {y}_{j}(k) = {y}_{j}^{*}, \text{ for } j=1,2,\ldots,M.
\end{align}
Analogously, the number of active energy consumers reaches its respective optimal value as in \eqref{eq:avg_consumer}.
\begin{align} \label{eq:avg_consumer}
\lim_{k\to \infty} {z}_{u}(k) = {z}_{u}^{*}, \text{ for } u=1,2,\ldots,U.
\end{align}
Thus, the community members achieve the minimum cost $\sum_{i=1}^{N} f_i(x_{i}^*) + \sum_{j=1}^{M} g_j(y_{j}^*) + \sum_{u=1}^{U} h_u(z_{u}^*)$ to the community over long-term averages.

\section{Algorithm}  \label{bin_imp} 
This section presents the developed distributed and stochastic algorithms for prosumers, consumers, and the central server. The prosumers with solar panels and the prosumers with wind turbines run their respective algorithms to share produced energy. Analogously, each consumer runs its algorithm to consume energy. At a time step, a prosumer is either active, producing energy, or not active; analogously, an energy consumer is either active, consuming energy, or not active. They decide to participate in energy production or consumption based on their response probabilities. 
We consider a central server that keeps track of the active prosumers (having solar panels and wind turbines) and consumers in the community. It updates and broadcasts feedback signals for prosumers and consumers.
Let us now define the parameters used in the solution. For $k \in \mathbb{N}$, let $\Theta^\textnormal{s}(k)$ denote the feedback signal for prosumers with solar panels and let $\Theta^\textnormal{w}(k)$ denote the feedback signal for the prosumers with wind turbines. Furthermore, let $\Theta^\textnormal{c}(k)$ denote the feedback signal for the energy consumers in the community. Let $\tau^\textnormal{s} \in (0,1)$ denote the step size for feedback signals for prosumers with solar panels. Analogously, let $\tau^\textnormal{w} \in (0,1)$ denote the step size for feedback signals for prosumers with wind turbines, and let $\tau^\textnormal{c} \in (0,1)$ denote the step size for feedback signals for consumers. We also call them as {\em gain parameters}. We choose suitable (fixed) values of the gain parameters for faster convergence.
When a prosumer with solar panels joins the smart energy community at time step $k$, $k \in \mathbb{N}$, it receives the parameter $\Theta^\textnormal{s}(k)$ from the central server. Analogously, prosumers with wind turbines receive $\Theta^\textnormal{w}(k)$ and the energy consumers receive $\Theta^\textnormal{c}(k)$ from the central server. The feedback signal for solar energy prosumers $\Theta^\textnormal{s}(k)$ depends on its value at the previous time step, the step size $\tau^\textnormal{s}$, the desired number of solar panels $C_\textnormal{s}$, and the total active solar energy prosumers in the community at the previous time step, for $k \in \mathbb{N}$.
The central server updates $\Theta^\textnormal{s}(k)$ according to \eqref{eq:Theta_s} at each time step $k$ and broadcasts it to all solar energy prosumers in the community:
\begin{align} \label{eq:Theta_s}
		\Theta^\textnormal{s}(k+1) &= \Theta^\textnormal{s}(k) - \frac{\tau^\textnormal{s}}{k+1} \left( \sum_{i=1}^{N} \xi_i(k) - C_\textnormal{s} \right).
\end{align}
Similarly, the central server updates $\Theta^\textnormal{w}(k)$ according to \eqref{eq:Theta_w} at each time step $k$, $k \in \mathbb{N}$, and broadcasts it to wind energy prosumers in the community. 
	\begin{align} \label{eq:Theta_w}
	    \Theta^\textnormal{w}(k+1) &= \Theta^\textnormal{w}(k) - \frac{\tau^\textnormal{w}}{k+1} \left( \sum_{j=1}^{M} \eta_j(k) - C_\textnormal{w} \right).
	\end{align}
Furthermore, the central server updates $\Theta^\textnormal{c}(k)$ according to \eqref{eq:Theta_c} at each time step $k$ and broadcasts it to all energy consumers in the community.
		
	\begin{align} \label{eq:Theta_c}
		&\Theta^\textnormal{c}(k+1) \nonumber \\&= \Theta^\textnormal{c}(k) - \frac{\tau^\textnormal{c}}{k+1} \left( \sum_{u=1}^{U} \zeta_u(k) - \sum_{i=1}^{N} \xi_i(k) - \sum_{j=1}^{M} \eta_j(k) \right).
  \end{align}
For a fixed $\tau^\textnormal{s} \in (0,1)$ and $k \in \mathbb{N}$, the step size $\frac{\tau^\textnormal{s}}{k+1}$ is decreasing; similarly, $\frac{\tau^\textnormal{w}}{k+1}$ and $\frac{\tau^\textnormal{c}}{k+1}$ are also decreasing. Interested readers can refer to \cite[Chapter 2]{Borkar2008} and \cite[Chapter 4]{Alam-PhDThesis2021} for an idea of the proof of convergence; it is an interesting future work.
After receiving the feedback signals, prosumers and consumers decide in a probabilistic way to be active or not. Let $\varphi^\textnormal{s}_{i}(x_i(k))$ denote the response probability for the prosumer with solar panels. The solar energy prosumer calculates the response probability $\varphi^\textnormal{s}_{i}(x_i(k))$ using its average number of times it was active up to time step $k$ and the derivative of its cost function, for $k \in \mathbb{N}$, as presented in \eqref{eq:prob_prod-sol}. The solar energy prosumer $i$ calculates the Bernoulli outcome at time step $k \in  \mathbb{N}$. Let $b_{i}(k)$ denote the Bernoulli outcome at time step $k$ for solar energy prosumer $i$, we have
\begin{align} \label{eq:bernoulli-solar}
b_{i}(k) = \left\{ \begin{array}{ll}
         1 & \mbox{with probability $\varphi^\textnormal{s}_{i}(x_i(k))$};\\
        0 & \mbox{with probability $1-\varphi^\textnormal{s}_{i}(x_i(k))$}.\end{array} \right.    
\end{align}
Based on the outcome, the solar energy prosumer decides whether to be active or not. If the value is $1$, then the solar prosumer is active; otherwise, it is not active. This process repeats over time. In a similar way, the wind energy prosumers and the energy consumers probabilistically decide to be active or not. Following this, the average number of times a prosumer was active converges to its optimal value. Let $b'_{j}(k)$ (cf. \eqref{eq:bernoulli-solar}) denote the Bernoulli outcome at time step $k$ for wind energy prosumer $j$ and $b''_{u}(k)$ (cf. \eqref{eq:bernoulli-solar}) denote the Bernoulli outcome at time step $k$ for energy consumer $u$, for $j=1,2,\ldots,M$, $u=1,2,\ldots,U$. The proposed algorithm for the central server is presented in Algorithm \ref{algoCU2}, the algorithm for prosumers with solar panels is presented in Algorithm \ref{algo_Producer_solar}, and the algorithm for prosumers with wind turbines is presented in Algorithm \ref{algo_Producer_wind}. Furthermore, the algorithm for consumers is presented in Algorithm \ref{algo_Consumer}.
	
\begin{algorithm}[!ht]  \SetAlgoLined Input:
		$C_\textnormal{s}$, $C_\textnormal{w}$, $\tau^\textnormal{s},\tau^\textnormal{w}$, $\tau^\textnormal{c}$, $\xi_{1}(k), \ldots, \xi_{N}(k)$, $\eta_{1}(k), \ldots, \eta_{M}(k)$, and $\zeta_{1}(k), \ldots, \zeta_{U}(k)$, for $k \in \mathbb{N}$.
		
		Output:
		$\Theta^\textnormal{s}(k+1), \Theta^\textnormal{w}(k+1)$, $\Theta^\textnormal{c}(k+1)$, for $k \in \mathbb{N}$.
		
		Initialization: $\Theta^\textnormal{s}(0)$, $\Theta^\textnormal{w}(0)$, $\Theta^\textnormal{c}(0) \in \mathbb{R}_+$ \;
		
		\ForEach{$k \in \mathbb{N}$}{
		
			calculate $\Theta^\textnormal{s}(k+1)$ as in \eqref{eq:Theta_s}	and broadcast it in the community\;
			calculate $\Theta^\textnormal{w}(k+1)$ as in \eqref{eq:Theta_w}	and broadcast it in the community\;
	       calculate $\Theta^\textnormal{c}(k+1)$ as in \eqref{eq:Theta_c} and broadcast it in the community;
		 }
		\caption{Algorithm of the central server.}
		\label{algoCU2}
	\end{algorithm}

\begin{algorithm}[!ht]  \SetAlgoLined Input:
	$\Theta^\textnormal{s}(k)$, for $k \in \mathbb{N}$.
	
	Output: $\xi_{i}(k+1)$, for $k \in \mathbb{N}$.
	
	Initialization: $\xi_{i}(0) \leftarrow 1$ and
	${x}_{i}(0) \leftarrow \xi_{i}(0)$.
	
	\ForEach{$k \in \mathbb{N} $}{
		
		
		calculate the response probability $\varphi^\textnormal{s}_{i}(x_i(k))$ as in \eqref{eq:prob_prod-sol}, obtain Bernoulli random variable
		$b_{i}(k)$ with probability of success $\varphi^\textnormal{s}_{i}(x_i(k))$ (see \eqref{eq:bernoulli-solar});
		
		\uIf{ $b_{i}(k) = 1$}{
			$\xi_{i}(k+1) \leftarrow 1$;
			
		}
		\Else{
				$\xi_{i}(k+1) \leftarrow 0$;}

}	
	\caption{Algorithm of prosumer $i$ with solar panels.}
	\label{algo_Producer_solar}
\end{algorithm}

\begin{algorithm}[!ht]  \SetAlgoLined Input:
	$\Theta^\textnormal{w}(k)$, for $k \in \mathbb{N}$.
	
	Output: $\eta_{j}(k+1)$, for $k \in \mathbb{N}$.
	
	Initialization: $\eta_{j}(0) \leftarrow 1$ and
	${y}_{j}(0) \leftarrow \eta_{j}(0)$.
	
	\ForEach{$k \in \mathbb{N} $}{
		
		
		calculate the response probability $\varphi^\textnormal{w}_{j}(y_j(k))$ as in \eqref{eq:prob_prod_wind}, obtain Bernoulli random variable
			$b'_{j}(k)$ with probability of success $\varphi^\textnormal{w}_{j}(y_j(k))$;
		
		
		\uIf{ $b'_{j}(k) = 1$}{
			$\eta_{j}(k+1) \leftarrow 1$;
			
		}
		\Else{
				$\eta_{j}(k+1) \leftarrow 0$;}

}	
	\caption{Algorithm of prosumer $j$ with wind turbines.}
	\label{algo_Producer_wind}
\end{algorithm}

\begin{algorithm}[!ht]  \SetAlgoLined Input:
	$\Theta^\textnormal{c}(k)$, for $k \in \mathbb{N}$.
	
	Output: $\zeta_{u}(k+1)$, for $k \in \mathbb{N}$.
	
	Initialization: $\zeta_{u}(0) \leftarrow 1$ and
	${z}_{u}(0) \leftarrow \zeta_{u}(0)$.
	
	\ForEach{$k \in \mathbb{N} $}{
		
		
		calculate the response probability $\varphi^\textnormal{c}_{u}(z_u(k))$ as in \eqref{eq:prob_cons}, obtain Bernoulli random variable
			$b''_{u}(k)$ with probability of success $\varphi^\textnormal{c}_{u}(z_u(k))$;
	
		\uIf{ $b''_{u}(k) = 1$}{
			$\zeta_{u}(k+1) \leftarrow 1$;
			
		}
		\Else{
				$\zeta_{u}(k+1) \leftarrow 0$;}

}	
	\caption{Algorithm of energy consumer $u$.}
	\label{algo_Consumer}
\end{algorithm}

	 After receiving the feedback signal $\Theta^\textnormal{s}(k)$ from the central server at time step $k$, prosumer $i$ with solar panels responds with probability $\varphi^\textnormal{s}_{i}(x_i(k))$ at the next time step, we define it as follows.
	\begin{align} \label{eq:prob_prod-sol}
		& \varphi^\textnormal{s}_{i}(x_i(k)) \triangleq  \Theta^\textnormal{s}(k) \frac{ x_{i}(k)}{f_i'(x_i(k))}.
	\end{align}
	Furthermore, after receiving the feedback signal $\Theta^\textnormal{w}(k)$ from the central server at time step $k$, prosumer $j$ with wind turbines responds with probability $\varphi^\textnormal{w}_{j}(y_j(k))$ at the next time step as follows.
	\begin{align} \label{eq:prob_prod_wind}
	  & \varphi^\textnormal{w}_{j}(y_j(k)) \triangleq  \Theta^\textnormal{w}(k) \frac{ y_{j}(k)}{g_j'(y_j(k))}.
	\end{align}
	Analogously, after receiving the feedback signal $\Theta^\textnormal{c}(k)$ from the central server at time step $k$, consumer $u$ responds with probability $\varphi^\textnormal{c}_{u}(z_u(k))$ at next time step as follows.
	\begin{align} \label{eq:prob_cons}
	& \varphi^\textnormal{c}_{u}(z_u(k)) \triangleq  \Theta^\textnormal{c}(k) \frac{ z_{u}(k)}{h_u'(z_u(k))}.
	\end{align}
	Note that $\Theta^\textnormal{s}(k)$, $\Theta^\textnormal{w}(k)$, and $\Theta^\textnormal{c}(k)$ are the feedback signals, they bound $\varphi^\textnormal{s}_{i}(x_i(k)), \varphi^\textnormal{w}_{j}(y_j(k)), \varphi^\textnormal{c}_{u}(z_u(k)) \in (0,1)$, respectively, for $i=1,2,\ldots,N$; $j=1,2,\ldots,M$; $u=1,2,\ldots,U$, and $k \in \mathbb{N}$. Following the above algorithms, the average number of times a prosumer with solar panels is active reaches its optimal value over time. Similarly, the average number of times a prosumer with wind turbines is active reaches its optimal value over time. And the average number of times a consumer is active reaches its optimal value over time; hence, the smart energy community achieves optimum social cost asymptotically. 
Nevertheless, we would like to clarify that though the prosumers and consumers keep their allocations, cost functions, and partial derivatives private; but, they could be inferred by adversary prosumers or consumers, especially when the number of participating members is small. Therefore, extending the algorithms to provide a privacy guarantee will be interesting.
%
%

\section{Numerical Results}
Consider that a smart energy community has $N$ prosumers with solar panels, $M$ prosumers with wind turbines, and $U$ energy consumers. The production and consumption of energy may incur costs to the prosumers and the consumers. The community members aim to minimize the overall cost to the community. Let $C_\textnormal{s}$ denote the total desired number of prosumers with solar panels, and $C_\textnormal{w}$ denote the total desired number of prosumers with wind turbines. In addition, we assume that each prosumer with solar panels has a private cost function $f_i(\cdot)$ which depends on the average of the number of times prosumer $i$ was active; recall that, we denote the average up to $k$ time steps by $x_{i}(k)$, for $i=1,2,\ldots,N$. Furthermore, each prosumer with wind turbines has a private cost function $g_j(\cdot)$ which depends on the average of the number of times prosumer $j$ with wind turbines was active; we denote the average until $k$ time steps by $y_{j}(k)$, for $j=1,2,\ldots,M$. Also, each consumer has a private cost function $h_u(\cdot)$ which depends on the average of number of times consumer $u$ was active, we denote the average up to $k$ time steps by $z_{u}(k)$, for $u = 1, 2, \ldots, U$. Moreover, we consider the central server aggregates the total number of active prosumers and consumers and broadcasts feedback signals in the community at each time step. Using the feedback signals, the derivative of the cost function, and the average number of times a member was active, it decides whether to be active in producing or consuming energy or not.

\subsection{Setup}
We chose $N = 100$ prosumers with solar panels, $M = 80$ prosumers with wind turbines, and $U = 160$ consumers in the energy community. Additionally, we chose capacities $C_\textnormal{s} = 50$ and $C_\textnormal{w} = 60$. Moreover, we chose the quadratic cost functions as in \cite{Fabio2019} to capture the costs. However, the simulation works as long as the cost functions are convex and increasing. Specifically, for solar energy prosumers, the following cost functions are used:  
	\begin{equation} \label{bin_func} 
	f_{i}(x_{i}) = a_{1i} x_{i} + b_{1i}x_{i}^2, \quad \textnormal{ for } i=1,2,\ldots,N.
\end{equation}
For wind energy prosumers, the following cost functions are used:  
	\begin{equation} \label{bin_func1} 
	g_{j}(y_{j}) = a_{2j} y_{j} + b_{2j} y_{j}^2, \quad \textnormal{ for } j=1,2,\ldots,M.
\end{equation}
And for energy consumers, we chose the following cost functions:  
\begin{equation}  \label{bin_func2}
	 h_{u}(z_{u}) = a_{3u} z_{u} + b_{3u} z_{u}^2, \quad \textnormal{ for } u=1,2,\ldots,U.
\end{equation}

To create randomized cost functions, we use random variables $a_{1i}, b_{1i}$, $a_{2j}$, $b_{2j}$, $a_{3u}$, and $b_{3u}$ that are uniformly distributed. We used Matlab to implement the algorithms. Furthermore, to compare the results, we solved the optimization problem \eqref{obj_fn1_b1} in a centralized way using CVX \cite{cvx2014}.

\subsection{Results}
\begin{figure*}[!ht]
	\centering
	\subfloat[]{
		\includegraphics[width=0.3\linewidth]{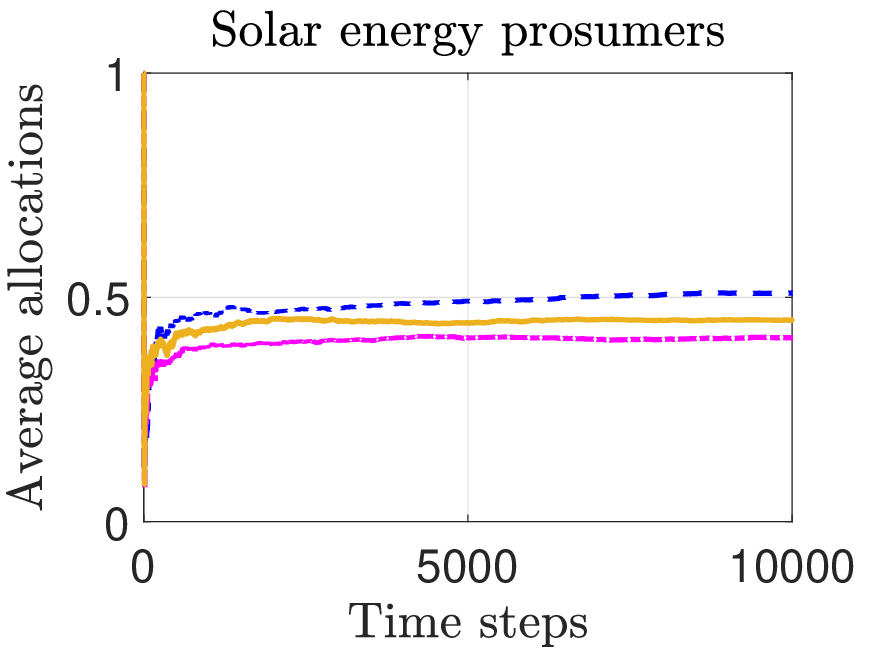}}
	\hfill
	\subfloat[]{ %
		\includegraphics[width=0.3\linewidth]{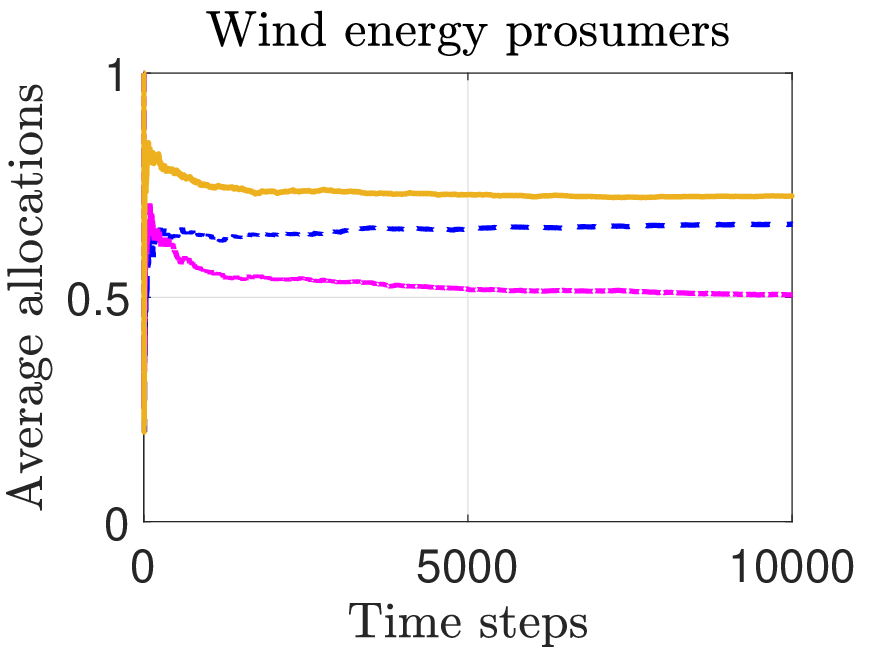}}
    \hfill
	\subfloat[]{ %
		\includegraphics[width=0.3\linewidth]{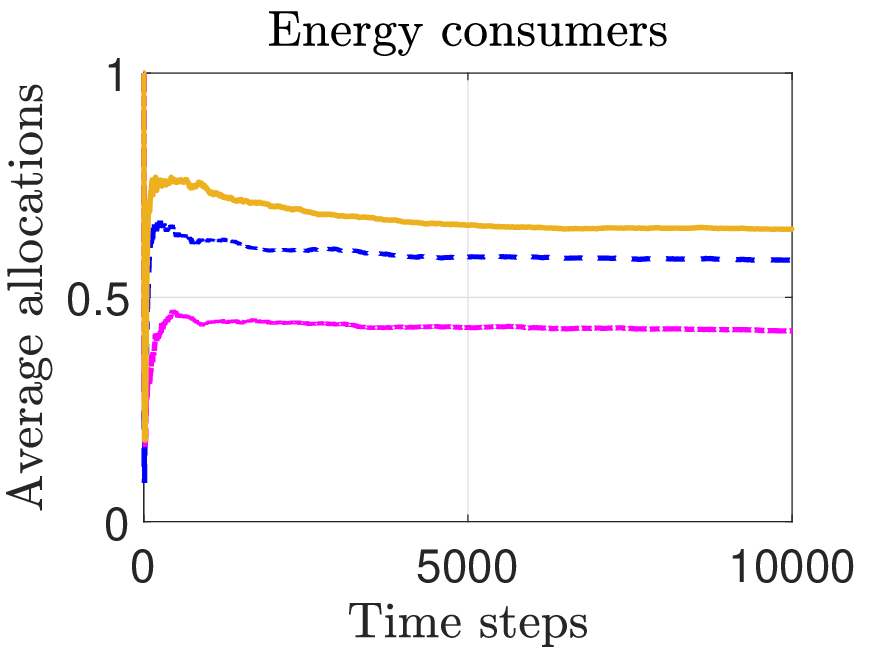}}
    \hfill
 \caption{(a) The evolution of the average number of times prosumers with solar panels are active. (b) The evolution of the average number of times prosumers with wind turbines are active. (c) The evolution of the average number of times consumers are active.} \label{fig1} 
\end{figure*}

\begin{figure*}[!ht]
	\centering
	\subfloat[]{
		\includegraphics[width=0.3\linewidth]{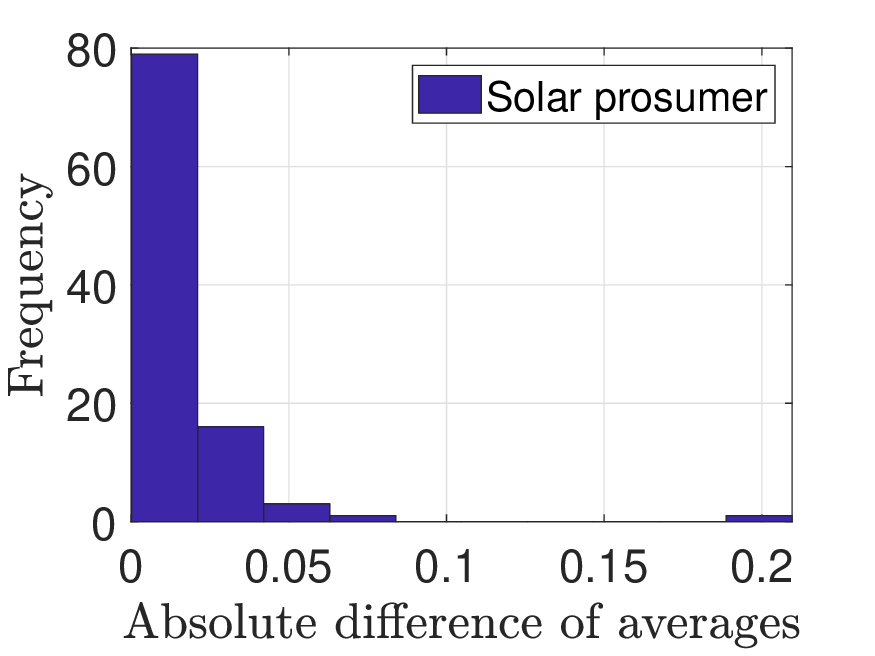}}
	\hfill
   \subfloat[]{
		\includegraphics[width=0.3\linewidth]{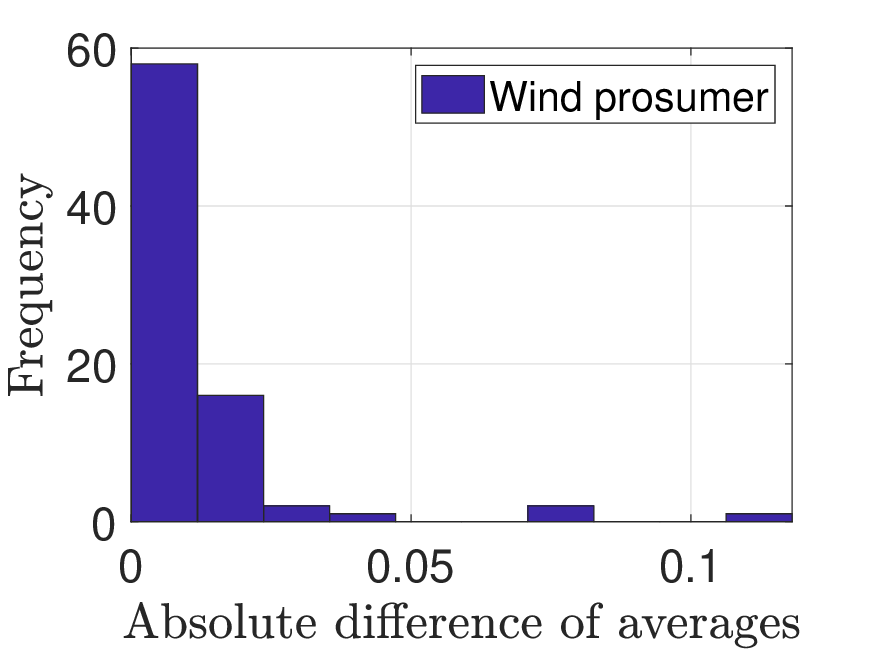}}
	\hfill
	\subfloat[]{
		\includegraphics[width=0.3\linewidth]{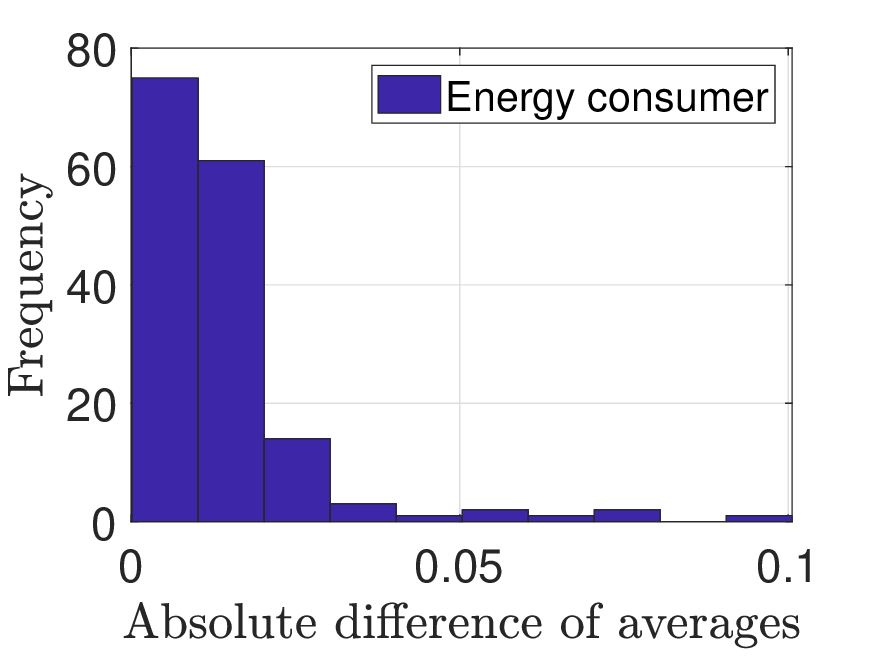}}
	\hfill
   \caption{(a) The histogram of absolute difference of the average number of active solar prosumers and the optimal value. (b) The histogram of the absolute difference between the average number of active wind prosumers and the optimal value. (c) The histogram of the absolute difference between the average number of active consumers and the optimal value (by the solver).} \label{fig2} 
\end{figure*}

We present the experimental results in this section. Figure \ref{fig1} shows the evolution of the average number of times prosumers with solar panels are active, prosumers with wind turbines are active, and the average number of times energy consumers are active. Furthermore, we observe that the time-averaged values converge over time. Suppose that $K$ denotes the last time step of the experiment. To check how close the average values for prosumers and consumers reach the respective optimal values, we plot the absolute difference between the average number of times prosumers and consumers are active up to time step $K$ and the respective optimal values.
In Figure \ref{fig2}, we plot $|x_i(K) - x_i^*|$, $|y_j(K) - y_j^*|$, and $|z_u(K) - z_u^*|$, for $i = 1, 2, \ldots, N$, $j = 1, 2, \ldots, M$, and $u = 1, 2, \ldots, U$, as histograms. Note that $x_i^*$, $y_j^*$, and $z_u^*$ are the optimal values obtained by the CVX solver, for $i=1,2,\ldots,N$, $j=1,2,\ldots,M$, and $u=1,2,\ldots,U$. Moreover, from Figure \ref{fig2}, we observe that the absolute difference between the averages and the optimal values is close to zero for most prosumers and consumers. 

\begin{figure}
	\centering
		\includegraphics[width=0.6\linewidth]{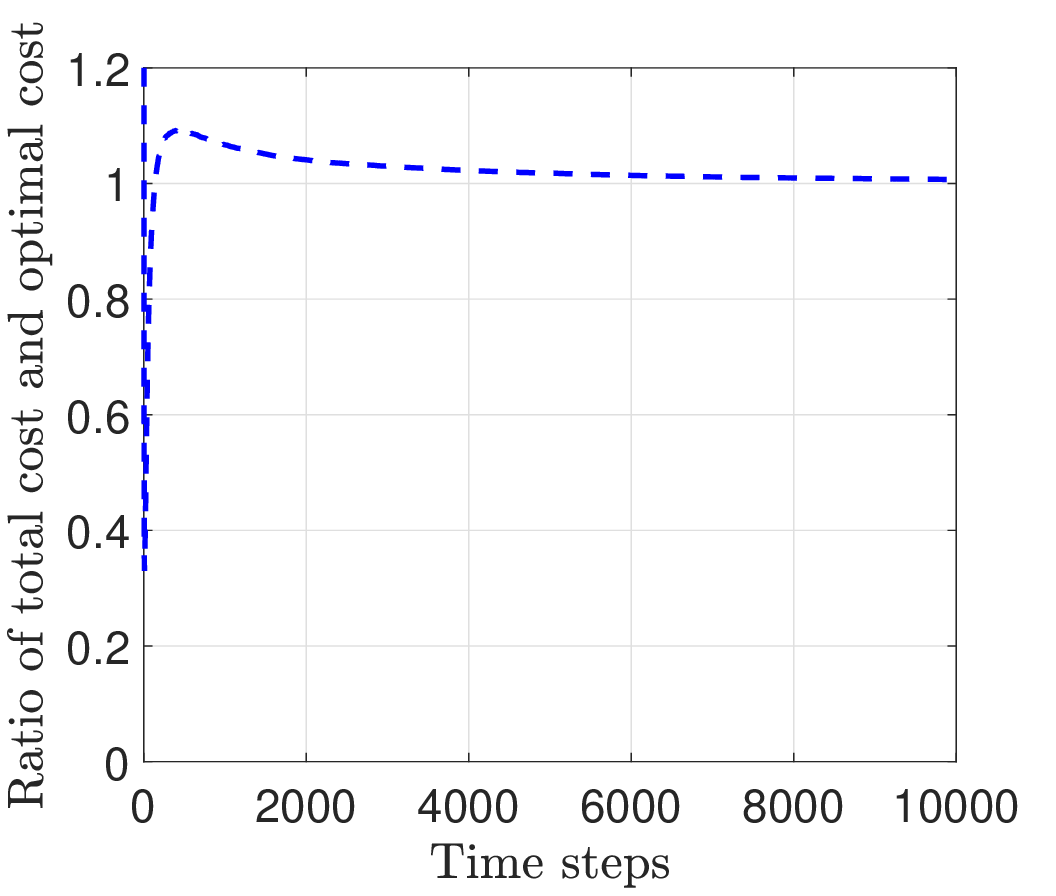}
	\hfill
   \caption{The evolution of the ratio of total cost obtained by the algorithms and the total optimal cost obtained by the solver.} \label{fig6}
 \end{figure}

For $\ell=1,2,\ldots,K$, Figure \ref{fig6} shows the evolution of the ratio of total cost $\sum_{i=1}^N f_i(x_i(\ell))  + \sum_{j=1}^M g_j(y_j(\ell)) + \sum_{u=1}^U h_u(z_u(\ell))$ and the total optimal cost by the solver $\sum_{i=1}^N f_i(x_i^*)  + \sum_{j=1}^M g_j(y_j^*) + \sum_{u=1}^U h_u(z_u^*)$. We observe that the ratio of total costs reaches close to $1$; that is, the total cost obtained by the algorithms and the total cost obtained by the solver approaches approximately the same value over time.  

\begin{figure*}[!ht]
	\centering
	\subfloat[]{
		\includegraphics[width=0.3\linewidth]{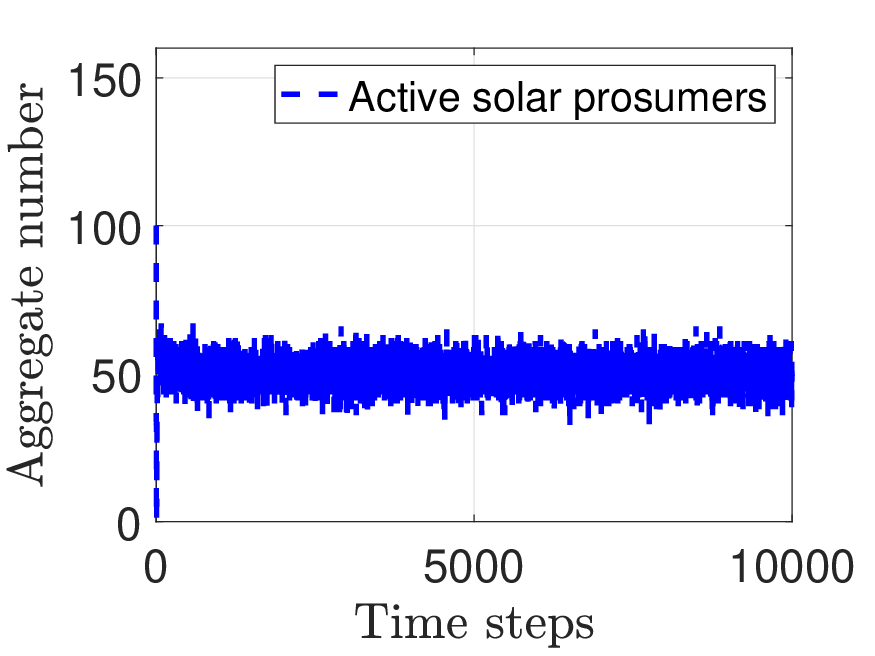}}
	\hfill
   \subfloat[]{
		\includegraphics[width=0.3\linewidth]{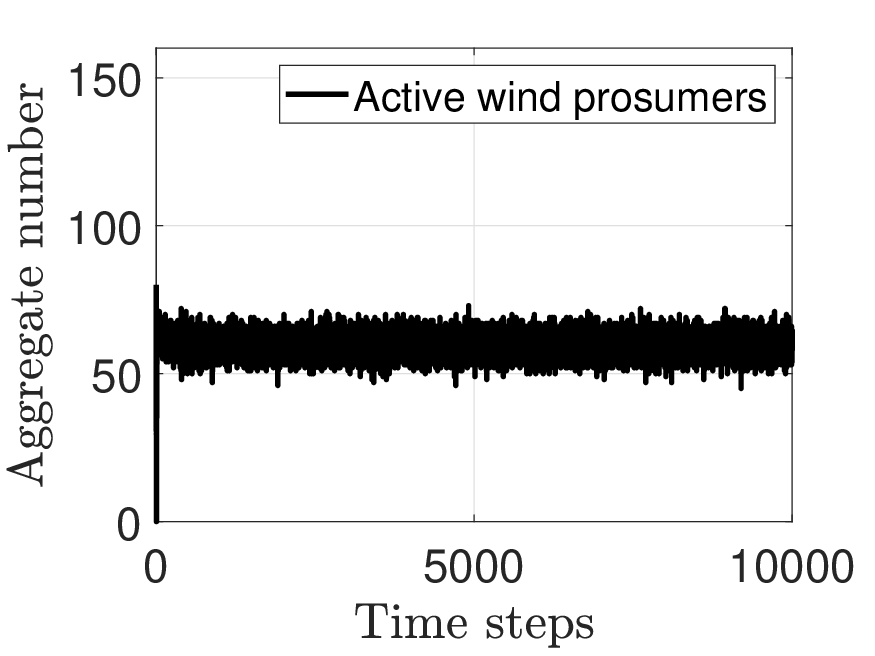}}
	\hfill
	\subfloat[]{
		\includegraphics[width=0.3\linewidth]{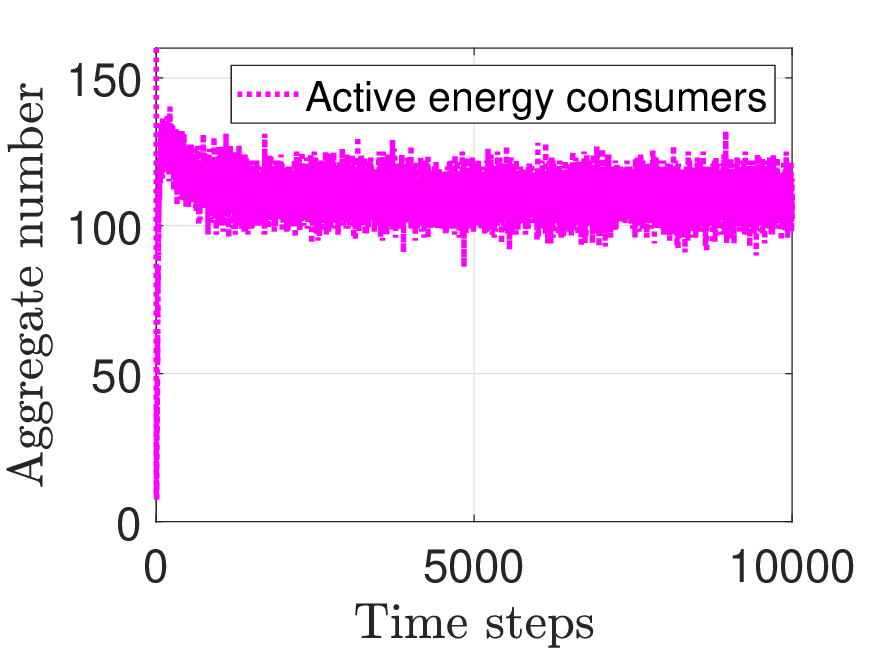}}
	\hfill
   \caption{(a) The evolution of the total number of active producers with solar panels, (b) the evolution of the total number of active producers with wind turbines, and (c) the evolution of the total number of active energy consumers.} \label{fig4} 
\end{figure*}

Finally, Figure \ref{fig4} illustrates the evolution of the total number of active prosumers with solar panels, the evolution of the total number of active prosumers with wind turbines, and the evolution of the total number of active energy consumers. We observe that the total number of active prosumers with solar panels is around its capacity, $C_\textnormal{s}$, and the total number of active prosumers with wind turbines is around its capacity $C_\textnormal{w}$. Furthermore, we observe that the total number of active consumers is around the sum of the capacities of prosumers with solar panels and the prosumers with wind turbines; that is, it fluctuates around $C_\textnormal{s} + C_\textnormal{w}$. The fluctuations are expected because of the randomized nature of the algorithms. 
 



\section{Conclusion}
We proposed distributed stochastic algorithms to regulate the number of prosumers and consumers in a smart energy community. We consider that the prosumers have heterogeneous energy sources---solar panels or wind turbines installed. In the model, the prosumers and consumers keep their information private such as its cost function, the derivative of the cost function, and when they were active. We consider a central server that keeps track of the aggregate number of active prosumers and consumers and sends feedback signals to the community members. Through experimental results, we show that the average number of times a prosumer and a consumer are active reaches their optimal values over time.
	
In future work, it will be interesting to extend this work to facilitate the consumers to track the energy source using blockchain technology. It may also enable the prosumers to track if the produced energy reaches the needy people when the prosumers are involved in philanthropic work. Furthermore, developing a blockchain-based system that does not require a central server and obtains optimal social value will be exciting. Last, it will be interesting to provide a theoretical guarantee on the convergence of average values and study the convergence rate of the algorithms. 

\bibliographystyle{IEEEtran}
\bibliography{P2PBib}

\end{document}